\documentclass[prb,showpacs,twocolumn,superscriptaddress]{revtex4}
\usepackage{hyperref,graphicx,dcolumn}
\usepackage{amsfonts, amsmath, amssymb, bm, bbm, mathtools, siunitx, physics}
\usepackage{color}
\usepackage{pdfsync}
\usepackage{blkarray}
\usepackage{multirow}
\usepackage{braket}
\usepackage{lipsum}

\newcommand{\epsdir}{.}

\newcommand{\myFig}[7]{ %
\begin{figure}[htb] 
\begin{center} 
\includegraphics[width=#1\columnwidth,height=#2\columnwidth,clip=true,keepaspectratio=#3,angle=#4]{\epsdir/#5}
\caption{#6} \vspace{-0.5cm} \label{#7} 
\end{center} \end{figure}}

\newcommand{\Norm}[1]{\| #1 \|}

\begin{document}
\title{{\it Ab initio} dynamical exchange interactions in frustrated anti-ferromagnets}
\author{Jacopo Simoni}\email[Contact email address: ]{simonij@tcd.ie}
\affiliation{School of Physics and CRANN Institute, Trinity College, Dublin 2, Ireland} 
\author{Maria Stamenova} 
\affiliation{School of Physics and CRANN Institute, Trinity College, Dublin 2, Ireland} 
\author{Stefano Sanvito} 
\affiliation{School of Physics and CRANN Institute, Trinity College, Dublin 2, Ireland} 

\begin{abstract}
The ultrafast response to an optical pulse excitation of the spin-spin exchange interaction in transition metal 
anti-ferromagnets is studied within the framework of the time-dependent spin-density functional theory. We 
propose a formulation for the purely dynamical exchange interaction, which is non-local in space, and it is 
derived starting from {\it ab initio} arguments. Then, we investigate the effect of the laser pulse on the onset 
of the dynamical process. It is found that we can distinguish two types of excitations, both activated immediately 
after the action of the laser pulse. While the first one can be associated to a Stoner-like excitation and involves 
the transfer of spin from one site to another, the second one is related to the ultrafast modification of a 
Heisenberg-like exchange interaction and can trigger the formation of spin waves in the first few hundred femtoseconds 
of the time evolution.           
\end{abstract}


\maketitle

\section{Introduction}
Density functional theory (DFT) has been the workhorse in material properties prediction from first principles 
for nearly half of a century. Among the many physical quantities that can be extracted from DFT, particularly
relevant for magnetism is the evaluation of the static Heisenberg exchange parameters~\cite{Kat00, Kat04, Loun10}.  
Their calculation has been closely related to and motivated by the problem of theoretically predicting the 
finite-temperature properties of magnetic systems. A possible approach consists in assuming that the magnetic 
excitations can be reasonably described by a Heisenberg-like Hamiltonian of the following form
\begin{equation}\label{Eq:1}
 H = -\frac{1}{2N}\sum_{m\neq n} J_{mn} \, \mathbf{S}_{m}\cdot \mathbf{S}_{n}\:,
\end{equation}
where $\mathbf{S}_{m}$ designates the spin vector associated to the site $m$, $J_{mn}$ is the exchange interaction 
between the spins at the two sites $m$ and $n$, and $N$ is the number of unit cells in the macroscopic system. 
If one considers a low energy excitation of the magnetic system described in terms of a spin spiral solution with 
wave vector $\mathbf{q}$ and polar angle $\theta$, the difference in total energy between this configuration, 
$E(\mathbf{q}, \theta)$, and the reference ferromagnetic one, $E(\mathbf{0}, \theta)$, will be in general related to 
the magnon frequency $\omega_{\mathbf{q}}$. In case of a single magnetic sublattice it can be shown 
that~\cite{EXCH3} 
\begin{equation}
 \omega_{\mathbf{q}} = 4 \frac{E(\mathbf{q}, \theta) - E(\mathbf{0}, \theta)}{M \sin^{2}\theta}\:,
\end{equation}
where $M$ is the magnitude of the onsite magnetization. Such frequency can be in turn related to the exchange 
parameter, $J(\mathbf{q})$, through the relation $\omega_{\mathbf{q}} = 2[J(\mathbf{0}) - J(\mathbf{q})]/M$. By employing 
the magnetic force theorem~\cite{MFT, MFT2, EXCH}, the difference in total energy between the two magnetic 
configurations can be related to the difference in the sum of the single particle energies calculated at the relevant 
spin densities. This allows one to estimate the bare exchange interaction directly from DFT results \cite{EXCH2}.

Exchange parameters can be also extracted from the dynamical linear response of the magnetic system to an external 
perturbation, that is usually expressed in terms of a small homogeneous magnetic field $\mathbf{b}^\mathrm{ext}(t)$. 
Exact susceptibilities can be, at least in principle, obtained from the time dependent extension of density functional theory 
(TDDFT). In Fourier space the linear response of the magnetization density~\cite{Liu89, TDDFT, Gross} writes
\begin{equation}
 \delta m_{-}(\mathbf{q}, \omega) = - \chi_{\pm}(\mathbf{q}, \omega) b^\mathrm{ext}_{-}(\mathbf{q}, \omega)\:,
\end{equation}
where the two functions $\delta m_{-}$ and $b^\mathrm{ext}_{-}$ are constructed through a linear combination of the $x$ and $y$ components of the
respective vectors in the form $f_{\pm}(\mathbf{q}, \omega) = (f_{x}\pm if_{y})(\mathbf{q},\omega)$, while $\chi_{\pm}(\mathbf{q}, \omega)$ represents the full spin-transverse 
susceptibility in Fourier space. 
The poles of $\chi_{\pm}(\mathbf{q}, \omega)$ define the excitation spectrum of the spin system, which in the 
zero-frequency limit returns the expression for the exchange coupling parameter of the effective Heisenberg 
Hamiltonian~\cite{Kat04}. In contrast, at higher frequencies the spin waves cannot be separated from the Stoner 
continuum.

The two methods just discussed both rest on an adiabatic assumption. Namely the timescales of the magnons and of 
the electronic motion differ enough to allow for the total energy differences between two magnetic configurations to be 
calculated within the framework of constrained non-collinear DFT. This, as it is well known, is designed to evaluate 
ground-state properties only. As a consequence neither the magnetic force theorem nor the calculation of the 
spin-transverse susceptibility are necessarily adequate to describe the out-of-equilibrium dynamics driven by very short 
(femtosecond scale) and strong laser pulses, when the electronic degrees of freedom cannot be averaged out.
One previous attempt to map the spin dynamics resulting from TDDFT simulations into the Heisenberg Hamiltonian
of Eq.~(\ref{Eq:1}) is based on a simple two-center molecule excited by very short and local in space magnetic 
fields~\cite{Stamen}. It was noticed that after the extinction of the pulse excitation the two atomic spins, deflected from the
collinear ground state to an angle $\phi$, display a precessional motion around the total spin axis with angular velocity given by 
$\omega = 4 J S \cos(\phi/2) / \hbar$, similarly to a pair of classical Heisenberg-coupled spins. This method was later extended 
to the study of the H - He - H magnetic molecule~\cite{Peral15}.
The external pulse used to excite the system in this work was only an instrumental one, acting as a small 
perturbation, which contributes very little as direct excitation to the electronic system. In this case the temporal evolution may be considered,
to a good degree of approximation, adiabatic.

However, the external fields cannot, in general, be treated as perturbations. This is certainly true for a class of ultrafast 
demagnetization phenomena \cite{Kim} discovered by Beaurepaire {\it et al} \cite{BEAU}, where an intense femtosecond laser 
pulse induces an abrupt loss of a large portion of the magnetization of a metallic film. There is little doubt that the 
exchange interaction plays a crucial r\^ole in the demagnetization observed at the femtosecond timescale and, in general, 
the spin dynamics in transition metal systems has always been explained within the framework of two different competing 
scenarios. In the first it is assumed that the main contribution to the spin dynamics can be attributed to collective magnonic 
excitations \cite{Carpene, EXCT2}, while in Reference~[\onlinecite{Sec13}] a new out-of-equilibrium spin-spin type of 
interaction was introduced starting from the Kadanoff-Baym formalism. The second scenario only considers the 
single-particle (Stoner) nature of the excitations in metals and recently it has been employed to justify ultrafast 
modifications of the exchange splitting driven by the external laser pulse \cite{Zhang15, dynEXCH}.     

In Ref.~\onlinecite{Elliot16} TDDFT calculations were employed to study the ultrafast magnetization dynamics in Heusler compounds, showing the important role
played by the spin currents in the process.
In this work we aim at introducing within the TDDFT framework the concept of {\it effective dynamical exchange interaction} (EDEI) and 
we will use such concept to analyze the laser-induced magnetization dynamics in anti-ferromagnetic metals directly 
in the time domain. The paper is organized as follows. In the next section we derive the fundamental equation of
motion for the magnetization density in TDDFT, and then we proceed to define the dynamical exchange splitting.  
Then we present our spin-dynamics results for metallic antiferromagnetic F\lowercase{e}M\lowercase{n} and finally we conclude.

\section{Methods}
When one neglects second-order contributions arising from the solution of the coupled Maxwell-Schr\"odinger 
system of equations, the dynamics is then governed by the following set of time-dependent Kohn-Sham (KS) 
equations
\begin{equation}\label{eq:TDKS}
 i\hbar\frac{d}{dt}\psi_{j}^{\mathrm{KS}}(\mathbf{r}, t) = H_{\mathrm{KS}}(\mathbf{r}, t)
 \psi_{j}^{\mathrm{KS}}(\mathbf{r}, t)\:.
\end{equation}
The KS Hamiltonian $H_\mathrm{KS}(\mathbf{r}, t)$ can be written by using the velocity gauge formulation and 
the minimal coupling substitution in the following form
\begin{align}
H_{\mathrm{KS}}(\mathbf{r}, t) & = \frac{1}{2m}\Big[ -i\hbar\nabla - \frac{q}{c}\mathbf{A}_{\mathrm{ext}}(t)\Big]^{2} + v_{\mathrm{s}}[n](\mathbf{r}, t) - \nonumber \\
& - \mu_{\mathrm{B}}\hat{\bm{\sigma}}\cdot\mathbf{B}_{\mathrm{s}}[n, \mathbf{m}](\mathbf{r}, t)\:,
\end{align}
where $v_{\mathrm{s}}(\mathbf{r}, t)$ represents the usual scalar KS potential, while
\begin{equation}\label{eq:Bs}
 \mathbf{B}_{\mathrm{s}}[n, \mathbf{m}](\mathbf{r}, t) = \mathbf{B}_{\mathrm{xc}}^{\mathrm{ALSDA}}[n, \mathbf{m}](\mathbf{r}, t) + \mathbf{B}_{\mathrm{ext}}(\mathbf{r}, t) \:.
\end{equation}
Here we have implied the use of the adiabatic local density approximation (ALSDA). The full non-interacting magnetic 
field $\mathbf{B}_{\mathrm{s}}(\mathbf{r}, t)$ is expressed as the sum of the external one and the exchange-correlation 
field, $\mathbf{B}_{\mathrm{xc}}(\mathbf{r}, t)=\delta E_\mathrm{xc}/\delta \mathbf{m}$, with $E_\mathrm{xc}$ being
the exchange-correlation energy. 

Starting from the set of equations~(\ref{eq:TDKS}) the following continuity equation for the spin density can be derived, 
\begin{align}\label{eq:spincon}
\frac{d\mathbf{m}(\mathbf{r},t)}{dt} & = -\nabla\cdot \mathbf{J}_{\mathrm{s}}(\mathbf{r}, t) + \frac{2\mu_{\mathrm{B}}}{\hbar}\mathbf{m}(\mathbf{r}, t)\times\mathbf{B}_{\mathrm{s}}(\mathbf{r},t) + \nonumber \\
& + \mathbf{T}_{\mathrm{SO}}(\mathbf{r}, t)\:,
\end{align}
where $\mathbf{T}_{\mathrm{SO}}(\mathbf{r}, t)$ defines the spin-orbit coupling contribution to the spin loss and 
$\mathbf{J}_{\mathrm{s}}(\mathbf{r}, t)$ is the non-interacting KS spin-current tensor. The KS magnetic field, 
$\mathbf{B}_{\mathrm{s}}(\mathbf{r},t)$, in absence of an external magnetic field simply reduces to the exchange-correlation 
contribution. In Refs.~[\onlinecite{Taka55, Antr1, Antr2, Antr3}] it was already pointed out that the spin current tensor
term can be rewritten in a different form through the prescription, 
$\nabla\cdot \mathbf{J}_{\mathrm{s}}(\mathbf{r}, t)=-\frac{2\mu_{B}}{\hbar} \mathbf{s}\times\mathbf{B}_{\mathrm{kin}}+\nabla\cdot[\mathbf{v}\mathbf{m}(\mathbf{r},t)]$. This expression, which introduces the so called {\it kinetic field}, $\mathbf{B}_{\mathrm{kin}}$, is 
however valid only in the single-particle case. For a many-particle system such reformulation of the divergence of 
the spin current leads to the following expression~\cite{me}
\begin{align}\label{eq:spincon2}
\frac{D}{Dt}\mathbf{m}(\mathbf{r},t) & = -\nabla\cdot\mathcal{D}(\mathbf{r},t) - \sum_{j\in\mathrm{occ.}}\mathbf{m}_j(\mathbf{r},t)\nabla\cdot\mathbf{v}_j(\mathbf{r},t) + \nonumber \\
& + \mu_{\mathrm{B}}\mathbf{m}(\mathbf{r},t)\times\mathbf{B}_{\mathrm{eff}}(\mathbf{r},t) + \mathbf{T}_{\mathrm{SO}}(\mathbf{r},t).
\end{align}
The term on the left-hand side of the equation is the material derivative, $\frac{D}{Dt}=\frac{d}{dt}+\mathbf{v}\cdot\nabla$, of the
magnetization density. On the right-hand side the term $\nabla\cdot\mathcal{D}(\mathbf{r},t)$ represents dissipation due to 
probability-current flow among different Kohn-Sham states,
\begin{align}
\mathcal{D}(\mathbf{r},t) & = -\sum_{j\in\mathrm{occ.}}\sum_{r\neq j\in\mathrm{occ.}}\mathcal{F}_{rj}\bigg[ \mathbf{J}_{\mathrm{s}}^{(j,r)}(\mathbf{r},t) - \nonumber \\
& - \mathbf{m}^{(r,j)}(\mathbf{r},t)\otimes\bigg(\mathbf{v}^{(j,r)}(\mathbf{r},t) + \frac{e}{mc}\mathbf{A}(\mathbf{r},t)\bigg)\bigg]\:,
\end{align}
with $\mathcal{F}_{rj}=\frac{\psi_r^{\mathrm{KS}\dagger}\psi_j^{\mathrm{KS}}}{n}$ and $n(\mathbf{r},t)$ being electron density 
of the system. The spin current field can be written as
\begin{equation}
\mathbf{J}_{\mathrm{s}}^{(j,r)}(\mathbf{r},t) = -\frac{i\hbar}{2m}\big[ \psi_j^{\mathrm{KS}\dagger}\hat{\boldsymbol{\sigma}}\nabla\psi_r^{\mathrm{KS}} - \nabla\psi_j^{\mathrm{KS}\dagger}\hat{\boldsymbol{\sigma}}\psi_r^{\mathrm{KS}}\big]\:,
\end{equation}
while the velocity field becomes
\begin{equation}
\mathbf{v}^{(j,r)}(\mathbf{r},t) = \frac{\hbar}{2mi}\frac{\psi_j^{\mathrm{KS}\dagger}\nabla\psi_r^{\mathrm{KS}} - \nabla\psi_j^{\mathrm{KS}\dagger}\psi_r^{\mathrm{KS}}}{\psi_j^{\mathrm{KS}\dagger}\psi_r^{\mathrm{KS}}} - \frac{e\mathbf{A}}{mc}\:.
\end{equation}
A second important term introduced in Eq.~(\ref{eq:spincon2}) has the form of an effective magnetic field acting 
on the magnetization density. This is
\begin{equation}\label{eq:beff}
\mathbf{B}_{\mathrm{eff}}(\mathbf{r},t) = \mathbf{B}_{\mathrm{s}}(\mathbf{r},t) + \frac{1}{\bar{\mathcal{F}}e}\bigg[ \frac{\nabla n(\mathbf{r},t)\cdot\nabla\mathbf{s}(\mathbf{r},t)}{n(\mathbf{r},t)} + \nabla^{2}\mathbf{s}(\mathbf{r},t)\bigg]\:,
\end{equation}
with $\bar{\mathcal{F}}=\frac{\expval{\psi_j^{\mathrm{KS}\dagger}\psi_j^{\mathrm{KS}}}_j}{n(\mathbf{r},t)}$.
The spin vector field $\mathbf{s}(\mathbf{r}, t)$ is defined through the relation 
$\mathbf{s}(\mathbf{r}, t)=\frac{\mathbf{m}(\mathbf{r}, t)}{n(\mathbf{r}, t)}$. The second term on the right-hand
side of Eq.~(\ref{eq:beff}) is the kinetic field
\begin{equation}\label{eq:bkin}
\mathbf{B}_{\mathrm{kin}}(\mathbf{r},t) = \frac{1}{\bar{\mathcal{F}}e}\bigg[ \frac{\nabla n(\mathbf{r},t)\cdot\nabla\mathbf{s}(\mathbf{r},t)}{n(\mathbf{r},t)} + \nabla^{2}\mathbf{s}(\mathbf{r},t)\bigg]\:.
\end{equation}
In Ref.~[\onlinecite{Taka55}] an expression analogous to that enclosed in the square brackets on the right-hand side 
of Eq.~(\ref{eq:bkin}) was identified as an effective dynamical exchange interaction responsible for possible spin-wave 
excitations in a magnetic system.

The charge continuity equation reads,
\begin{equation}\label{eq:ncont}
\frac{D}{Dt}n(\mathbf{r}, t) = -n(\mathbf{r}, t)\nabla\cdot\mathbf{v}(\mathbf{r}, t)\:,
\end{equation}
and it is valid for the density of every single KS state. It may also be rewritten in the form 
$\frac{d}{dt}n(\mathbf{r},t)=-\nabla\cdot[n\mathbf{v}]$, with 
$\mathbf{v}(\mathbf{r},t) = \frac{\mathbf{j}_{\mathrm{p}}(\mathbf{r},t)}{n(\mathbf{r},t)} - \frac{e}{mc}\mathbf{A}(\mathbf{r},t)$ 
and $\mathbf{j}_{\mathrm{p}}(\mathbf{r},t)$ being the paramagnetic current of the non-interacting system. 
We have determined that the electron density variation during the action of the laser pulse can be considered, 
in our calculations, much smaller than the temporal variation of the magnetization density (see for instance Fig.~\ref{fig:02}(a)). By considering an approximately 
homogeneous electron density, $\bar{n}(t)$, in the vicinity of the atoms we have $\dot{\bar{n}} = -\bar{n}(t)\nabla\cdot\mathbf{v}$. 
Thus the small value of $\dot{\bar{n}}$ compared to $\dot{\mathbf{m}}(\mathbf{r},t)$ suggests that, in first approximation, the 
velocity field $\mathbf{v}(\mathbf{r},t)$ can be safely neglected from our discussion on the spin dynamics. The spin-orbit coupling 
contribution to the dynamics can also be neglected because much weaker than the other terms appearing in Eq.~(\ref{eq:spincon2}).

In conclusion we are left with the following simplified equation of motion for the magnetization,
\begin{equation}\label{eq:spincon3}
\frac{d}{dt}\mathbf{m}(\mathbf{r}, t) = -\nabla\cdot\mathcal{D}(\mathbf{r},t) + \mu_{B}\mathbf{m}(\mathbf{r},t)\times\mathbf{B}_{\mathrm{eff}}(\mathbf{r},t)\:.
\end{equation}
The effective field $\mathbf{B}_{\mathrm{eff}}(\mathbf{r},t) = \mathbf{B}_{\mathrm{s}}(\mathbf{r},t)+\mathbf{B}_{\mathrm{kin}}(\mathbf{r},t)$ 
is not necessarily parallel to the magnetization $\mathbf{m}(\mathbf{r}, t)$ at every point in space, hence it can produce an effective 
contribution to the dynamics of the magnetization vector.

In the absence of an external magnetic field, $\mathbf{B}_{\mathrm{s}}=\mathbf{B}_{\mathrm{xc}}$, and the properties 
of the two components of $\mathbf{B}_{\mathrm{eff}}$, within the ALDA, have been already described elsewhere \cite{spindyn, me}. 
Here our aim is at extracting a more conventional physical interpretation of the role of $\mathbf{B}_{\mathrm{xc}}$ and $\mathbf{B}_{\mathrm{kin}}$ 
during the evolution of the system far away from equilibrium and their relation to established spin dynamics models. We start this analysis by noting that the expression for
$\mathbf{B}_{\mathrm{xc}}(\mathbf{r},t)=\mathbf{B}_{\mathrm{x}}(\mathbf{r},t)+\mathbf{B}_{\mathrm{c}}(\mathbf{r},t)$ is local in space
within the ALDA. In fact $\mathbf{B}_{\mathrm{xc}}(\mathbf{r},t)$ depends uniquely on the value of density and magnetization at the 
given point. 
%
%
The same argument cannot be used for the kinetic field. In fact 
the expression~(\ref{eq:bkin}) does not depend explicitly on the spin vector $\mathbf{s}(\mathbf{r}, t)$, but on its gradient $\nabla\mathbf{s}(\mathbf{r},t)$. 
A consequence of such property of $\mathbf{B}_{\mathrm{kin}}$ is that at every point in space the value of the field depends not only on 
the value of the magnetization at that particular point, but also on the value of the spin vector in its vicinity. 

In the next Section \ref{sect:2} we investigate the possibility to rewrite $\mathbf{B}_{\mathrm{kin}}(\mathbf{r},t)$ in a form 
where its dependence on the spin vector becomes explicit and the local and semi-local contributions of the spin gradient are 
separated. In Section \ref{sect:3} we look at the ultrafast magnetization dynamics of the frustrated anti-ferromagnet 
F\lowercase{e}M\lowercase{n} by analyzing the contribution of the different magnetic excitations and in particular by focussing 
on the role of the EDEI at ultrafast time scales. Finally in Section \ref{sect:4} we conclude.

\section{The dynamical exchange interaction} \label{sect:2}

We start by rewriting the kinetic field, $\mathbf{B}_{\mathrm{kin}}$, introduced in Eq.~(\ref{eq:bkin}). This object could be 
thought as a vector field defined on a three-dimensional space spanned by the spin vector components, namely
\begin{equation}
 \mathbf{B}_{\mathrm{kin}}(\mathbf{r},t) = \big( F[\mathbf{f}_{1}, n], F[\mathbf{f}_{2}, n], F[\mathbf{f}_{3}, n]\big)\:,
\end{equation}
where we have introduced the scalar functional,
\begin{equation}\label{funct}
 F[\mathbf{f}, n] = \frac{\boldsymbol{\nabla}n}{n}\cdot\mathbf{f} + \boldsymbol{\nabla}\cdot\mathbf{f}\:,
\end{equation}
and $\mathbf{f}(\mathbf{r}, t)$ represents a generic differentiable vector field in $\mathbb{R}^3$ such that 
\begin{equation}
f_{i}(\mathbf{r},t) \in C^{1}(\mathbb{R}^{3}\times[0, +\infty))\:\:\:\: \textrm{ for i} = 1, 2, 3 \:.
\end{equation}
Suppose now that we want to evaluate the functional in Eq.~(\ref{funct}) at a certain point in space 
$x_{0}\in \mathbb{R}^{3}\times[0, +\infty)$. The function $\mathbf{f}(x)$ may be then separated into 
a local and a non-local part around $x_{0}$ as follows
\begin{equation}\label{eq:decomp}
\mathbf{f}(x_{0}) = \mathbf{h}(x_{0}) + \int\dd[3]{x} \boldsymbol{\epsilon}(x_{0}, x) \eta(x)\:,
\end{equation}
where $\boldsymbol{\epsilon}(x_{0}, x)$ and $\eta(x)$ are respectively a non-local vector field and a scalar field.
Hence, at $x_{0}$ one can write
\begin{align}
F[\mathbf{f}, n](x_{0}) = & \frac{\boldsymbol{\nabla}_{0}n}{n(x_{0})}\cdot\Big[ \mathbf{h}(x_{0}) + \int \dd[3]{x}\boldsymbol{\epsilon}(x_{0}, x)\eta(x)\Big] + \nonumber \\
& + \boldsymbol{\nabla}_{0}\cdot\Big[ \mathbf{h}(x_{0}) + \int \dd[3]{x} \boldsymbol{\epsilon}(x_{0}, x)\eta(x) \Big]\:.
\end{align}
By separating in the previous expression the local from the non-local contribution we have
\begin{align}
& F[\mathbf{f}, n](x_{0}) = \frac{\boldsymbol{\nabla}_{0}n}{n(x_{0})}\cdot\mathbf{h}(x_{0}) + \boldsymbol{\nabla}_{0}\cdot\mathbf{h}(x_{0}) + \nonumber \\
& + \int \dd[3]{x} \Big[ \frac{\boldsymbol{\nabla}_{0}n}{n(x_{0})}\cdot\boldsymbol{\epsilon}(x_{0}, x) + \boldsymbol{\nabla}_{0}\cdot\boldsymbol{\epsilon}(x_{0}, x)\Big] \eta(x)\:,
\end{align}
where we identify a local field,
\begin{equation}
B_{\mathrm{local}}(x_{0}) = \frac{\boldsymbol{\nabla}_{0}n\cdot\mathbf{h}(x_0)}{n(x_{0})} + \boldsymbol{\nabla}_{0}\cdot\mathbf{h}(x_{0})\:,
\end{equation}
and an effective non-local field,
\begin{equation}
J(x, x_{0}) = \frac{\boldsymbol{\nabla}_{0}n}{n(x_{0})}\cdot\boldsymbol{\epsilon}(x_{0}, x) + \boldsymbol{\nabla}_{0}\cdot\boldsymbol{\epsilon}(x_{0}, x)\:.
\end{equation}
We now need a proper definition for the non-local vector field, $\boldsymbol{\epsilon}(x_{0}, x)$. This definition depends 
on the choice of the scalar field, $\eta(x)$, in Eq.~(\ref{eq:decomp}). Here we substitute $\eta(x)$ with a given component 
$e_{i}(x)$ of the unitary magnetization vector
\begin{equation}
\mathbf{f}_{i}(x_{0}) = \mathbf{h}_{i}(x_{0}) + \int\dd[3]{x} \boldsymbol{\epsilon}_{i}(x_{0}, x) e_{i}(x)\:.
\end{equation}
By taking the average of the unitary magnetization component $e_{i}(x)$ over the integration region we can approximate 
the integral as follows
\begin{equation}
\mathbf{f}_{i}(x_{0}) = \mathbf{h}_{i}(x_{0}) + \boldsymbol{\bar{\epsilon}}_{i}(x_{0}) \bar{e}_{i}\:,
\end{equation}
where $\mathbf{f}_{i}(x_{0})$ has been separated into two components, the first orthogonal to the spin direction, 
$\bar{e}_{i}$, and the second parallel to it. In this form $\boldsymbol{\bar{\epsilon}}_{i}(x_{0})$ defines simply the 
projection of the vector $\mathbf{f}_{i}(x_{0})$ along the direction $\bar{e}_{i}$ in spin space
\begin{equation}
\boldsymbol{\bar{\epsilon}}_{i}(x_{0}) = \langle \mathbf{f}_{i}(x_{0}), \bar{e}_{i} \rangle\:.
\end{equation}
By substituting $\mathbf{f}_{i}(x_{0})$ with the gradient of the spin vector, $\nabla s_{i}(x_{0})$, we are now in 
the position to separate the local from the non-local component of the kinetic field, $\mathbf{B}_{\mathrm{kin}}(\mathbf{r},t)$, 
of Eq.~(\ref{eq:bkin}). From the linearity of the functional $F[\mathbf{f}, n]$ in the $\mathbf{f}$ variable,
\begin{equation}
F[\mathbf{h}_{i}+\boldsymbol{\bar{\epsilon}}_{i} \bar{e}_{i}, n](x_{0}) = F[\mathbf{h}_{i}, n](x_{0}) + F[\boldsymbol{\bar{\epsilon}}_{i}, n](x_{0}) \bar{e}_{i}.
\end{equation}
The first term gives rise to an effective local field of the following form,
\begin{equation}
\hat{B}_{\mathrm{local}}(\mathbf{r}, t) = \big( F[\hat{\mathbf{h}}_{1}, n](\mathbf{r},t), F[\hat{\mathbf{h}}_{2}, n](\mathbf{r},t), F[\hat{\mathbf{h}}_{3}, n](\mathbf{r},t)\big)\:,
\end{equation}
while the second term on the right-hand side can be rewritten in a way that displays an explicit dependence on the spin 
vector, $\hat{\bar{s}}$, thus generating a new effective mean field object \footnote{Here $\hat{\bar{s}}$ indicates that we 
are considering a simple vector in spin space, the number of points used in the average depends on the definition of the 
gradient over the grid, usually $4$ points for every direction with a total of $12$.}
\begin{equation}
\hat{B}_{\mathrm{mf}}(\mathbf{r}, t) = \sum_{i=1}^{3}F[\boldsymbol{\bar{\epsilon}}_{i}, n](\mathbf{r}, t) \langle \hat{\bar{e}}_{i}, \hat{\bar{s}} \rangle \hat{\bar{s}}\:.
\end{equation}
The kinetic magnetic energy can be, therefore, finally separated into two contributions
\begin{equation}
E_{\mathrm{kin}}[n, \hat{s}] = \int\dd[3]{r} \hat{s}(\mathbf{r},t)\cdot\big[ \hat{B}_{\mathrm{local}}(\mathbf{r},t) + \hat{B}_{\mathrm{mf}}(\mathbf{r},t)\big]\:.
\end{equation}
By summing up the non-interacting part of the energy with the interacting one dominated by the exchange-correlation 
potential, we obtain
\begin{align}\label{eq:emag}
E_{\mathrm{kin+xc}}[n,\hat{s}] & = \int\dd[3]{r} \hat{s}(\mathbf{r},t)\cdot\big[ \hat{B}_{\mathrm{local}}(\mathbf{r},t) + \hat{B}_{\mathrm{xc}}(\mathbf{r},t)\big] + \nonumber \\
& + \int\dd[3]{r} \hat{s}(\mathbf{r},t)\cdot\hat{B}_{\mathrm{mf}}(\mathbf{r},t)\:.
\end{align}
While the first term on the right-hand side of Eq.~(\ref{eq:emag}) represents a dynamical Stoner-like field, the nature 
of the second term, due to its spatial non-locality, is completely different and resembles the form of a Heisenberg exchange 
with mean field energy
\begin{equation}\label{eq:spininter}
E_{\mathrm{mf}}[n, \hat{s}] = \sum_{i=1}^{3}\int \dd[3]{r} \hat{s}(\mathbf{r},t)\cdot F[\boldsymbol{\bar{\epsilon}}_{i}, n](\mathbf{r},t) \langle\hat{\bar{e}}_{i}, \hat{\bar{s}}\rangle \hat{\bar{s}}\:.
\end{equation}
From the previous expression we can finally identify an EDEI
\begin{equation}\label{eq:jcoupl}
J_{\mathrm{mf}}(\mathbf{r}, t) = \sum_{i=1}^{3} F[\boldsymbol{\bar{\epsilon}}_{i}, n](\mathbf{r},t) \langle\hat{\bar{e}}_{i}, \hat{\bar{s}}\rangle\:.
\end{equation}

\section{Ultrafast spin dynamics in F\lowercase{e}M\lowercase{n}}\label{sect:3}
\myFig{1}{1}{true}{0}{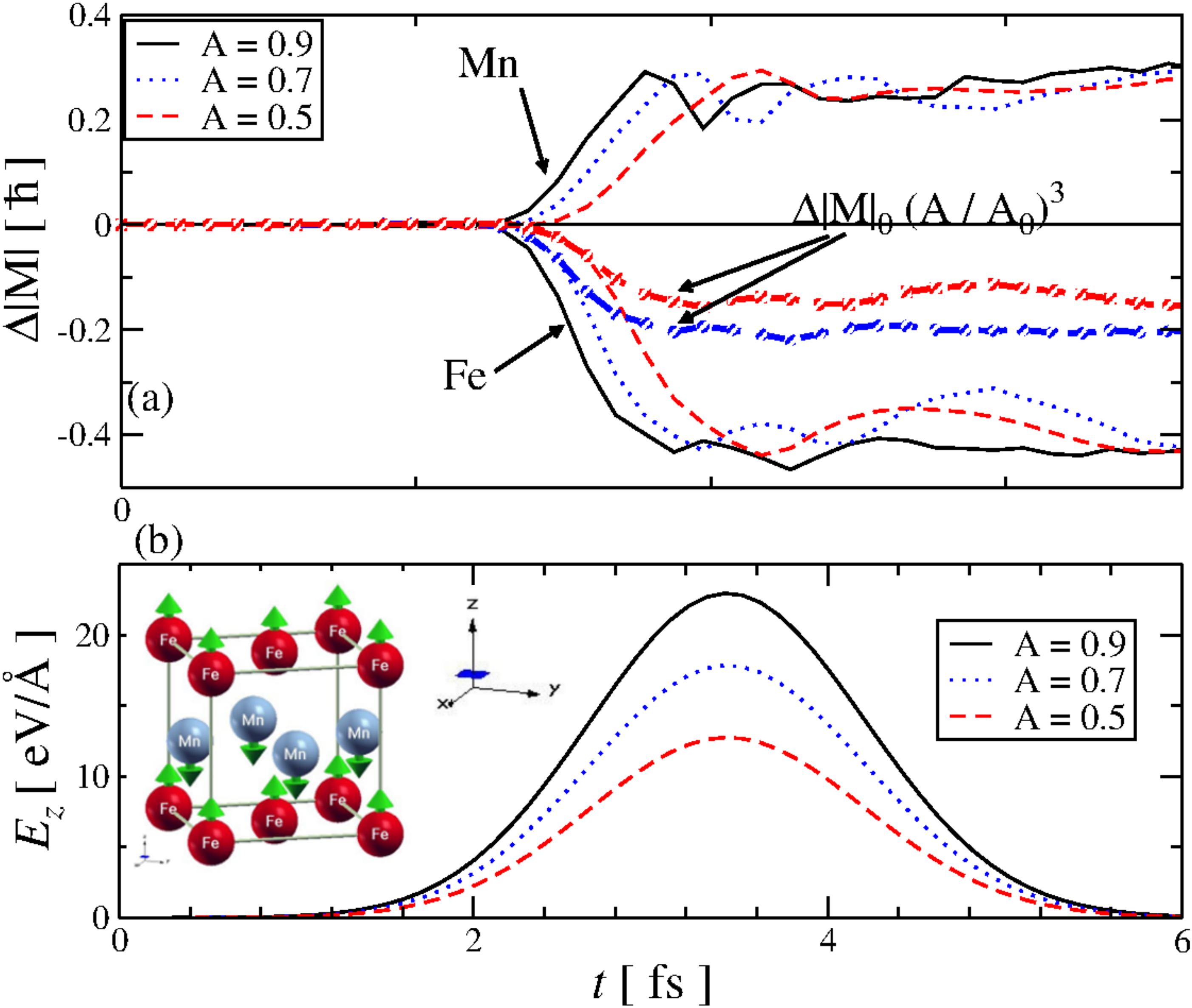}{(Color online) Time-averaged observables evaluated in spheres of radius $\SI{0.5}{\angstrom}$ 
around each atom: (a) the averaged temporal variation of the spin density module 
$\abs{\Delta\mathbf{S}^{\mathrm{N}}(t)}=\int_{\mathcal{S}_{\mathrm{N}}^{\mathrm{R}}}\dd[3]{r}(\abs{\mathbf{s}(\mathbf{r},t)}-\abs{\mathbf{s}(\mathbf{r},0)})$ 
under various laser pulses for the two F\lowercase{e} and M\lowercase{n} sites; (b) the different laser pulses employed in the 
dynamical simulations with shape $E_{\mathrm{z}}(t)=A\cdot\sin(0.1\cdot t)\cdot\exp(-(t-5)^{2} / 3)$ for t in fs. The amplitudes $A$ are in 
units of eV\ /$\mathring{A}$ (the black arrow in the inset indicates the electric field polarization direction). }{fig:01}
In order to analyze how the quantities previously defined evolve dynamically in a real magnetic system under the 
action of an external electric pulse, we look at bulk F\lowercase{e}M\lowercase{n}. The ground state properties of 
this material have been already studied in the past~\cite{femn1, femn2}, even though there is no full consensus 
on the magnetic structure of the ground-state, since the various theoretical results often vary with the method and 
approximation employed. Here we consider the anti-ferromagnetic ground-state in its fcc phase with lattice constant 
$a=\SI{3.7}{\angstrom}$ [see inset in Fig.~\ref{fig:01}(b)]. This structure represents the starting point of our dynamical 
evolution. We use the ALDA \cite{alda} exchange-correlation functional with the Perdew and Wang \cite{Wang92} 
parametrization as implemented in the {\sc Octopus} code \cite{Octop1}. The ground state is characterized by two 
localized magnetic moments over the F\lowercase{e} and M\lowercase{n} sites with a magnitude 
$\abs{\mathbf{S}}\simeq 0.57~\hbar$ computed by integrating the spin density within atom centered spheres of radius $\SI{0.5}{\angstrom}$. 
The amount of non collinearity is not negligible but the ratio among $m_{\mathrm{z}}$ and $m_{\mathrm{x}}$ 
(or $m_{\mathrm{y}}$) is always approximately $4$ and it has the tendency to increase with the distance from the atom. 
The $z$ component of the magnetization vector is thus locally dominant, even if over the entire simulation box it is 
approximately zero due to the overall anti-ferromagnetic nature of the system.

In all our calculations the system is perturbed from the initial equilibrium ground state by applying intense, spatially 
homogeneous, electric pulses, with duration typically between $7$~fs and $10$~fs. The pseudopotentials for both 
F\lowercase{e} and M\lowercase{n} employed in the calculations are fully relativistic and norm-conserving. They are 
generated using a multi-reference pseudo-potential (MRPP) scheme~\cite{mrpp} at the level implemented in 
{\sc APE}~\cite{APE, APE2}, which evolves the valence states and the semi-core states simultaneously.

In order to analyze the spin dynamics in an anti-ferromagnetic material we need to partition the spin density so to isolate 
the magnetic moments and the electronic charges associated with each atomic site $N$ in the unit cell. The simplest choice 
consists in integrating the densities inside a sphere $\mathcal{S}_{\mathrm{N}}^{\mathrm{R}}$ of radius $R$ centered on 
the atomic site $N$. Thus the local spin and charge densities read respectively
\begin{equation}\label{eq:Mloc}
\mathbf{S}^{\mathrm{N}}(t) = \int_{\mathcal{S}_{\mathrm{N}}^{\mathrm{R}}} \dd[3]{r}\mathbf{s}(\mathbf{r},t)\:,\:\:\: 
\quad Q^{\mathrm{N}}(t) = \int_{\mathcal{S}_{\mathrm{N}}^{\mathrm{R}}}\dd[3]{r}\; n(\mathbf{r},t)\:.
\end{equation}
In Fig.~\ref{fig:01}(a) we show the demagnetization observed around the F\lowercase{e} and the M\lowercase{n} sites 
in the first $20$~fs under different laser pulses, all polarized along the $z$ direction but with different amplitudes. The 
demagnetization process is quite pronounced since, in all the cases, each atom loses around $60\%$ of the initial magnetization 
almost immediately after the action of the pulse and then it stabilizes around a different value of the magnetization vector. 
The demagnetization rate, instead, differs in the three cases. In particular we observe from Fig.~\ref{fig:01}(a) an initial decay rate proportional to the 
the cube of the laser field amplitude
\begin{equation}
 \abs{\mathbf{S}(t_{\mathrm{in}}+\delta t)} - \abs{\mathbf{S}(t_{\mathrm{in}})} \sim - A^3\:,
\end{equation}
where $t_{\mathrm{in}}$ is the initial time at which the laser pulse is applied, $\delta t$ is a small time step, while $A$ represents the 
amplitude of the applied laser pulse in $eV /\mathring{A}$. Hence the demagnetization rate increases substantially for larger 
excitation amplitudes. At the same time the overall magnetization loss for longer times following the laser pulse does not change 
significantly.

The observed magnetization dynamics, localized in the vicinity of the two atomic sites, suggests the existence of a spin density 
transfer mechanism among different occupied and unoccupied Kohn-Sham states. Consider now the magnetization continuity 
equation~(\ref{eq:spincon3}), where the velocity field has been, in first approximation, neglected. We have that the dissipative 
term, $-\nabla\cdot\mathcal{D}(\mathbf{r},t)$, on the right-hand of the expression, is the one driving the entire dynamics during 
the action of the laser pulse. This influences also the other field, $\mathbf{B}_{\mathrm{eff}}(\mathbf{r},t)$, that is modified by the 
local changes in the spin density gradient.

After having neglected in Eq.~(\ref{eq:spincon3}) the exchange-correlation magnetic field, $\mathbf{B}_{\mathrm{xc}}(\mathbf{r}, t)$, 
which does not contribute to the dynamics in the ALDA, but represents only an energy barrier between the spin-up and spin-down states, 
we are left with the equation
\begin{equation}\label{eq:spincon4}
\frac{d}{dt}\mathbf{m}(\mathbf{r}, t) = -\nabla\cdot\mathcal{D}(\mathbf{r},t) + \mu_{B}\mathbf{m}(\mathbf{r},t)\times\mathbf{B}_{\mathrm{kin}}(\mathbf{r},t)\:.
\end{equation}
Here we finally distinguish two contributions to the spin dynamics. The first one on the right-hand side of Eq.~(\ref{eq:spincon4}) 
represents a measure of the spin dissipation due to the internal charge currents flowing between the different Kohn-Sham states.
This term is, by construction, responsible for effective Stoner-like excitations in real time. In fact, Stoner excitations 
are due, by definition, to terms in the Hamiltonian of the form $\hat{c}^{\dagger}_{k,A,\uparrow}\hat{c}_{k,B,\downarrow}$, where $A$ and $B$ 
label the atomic site. These excitations are local in momentum space and non-local in real space and correspond to 
inter-site electronic excitations, which are included into the previous spin dissipation term.

The second contribution to the dynamics is due to the torque exerted by the kinetic field, $\mathbf{B}_{\mathrm{kin}}(\mathbf{r}, t)$, on the magnetization vector. Due to its dependence on the 
gradients of the electron density and magnetization $\mathbf{B}_{\mathrm{kin}}$ drastically changes during the action of the 
electric pulse. In Fig.~\ref{fig:02}(b) we compare the evolution of the $z$ component of the magnetization with its transverse component 
$S_{\mathrm{xy}}^{\mathrm{N}}(t)=\int_{\mathcal{S}_{\mathrm{R}}^{\mathrm{N}}}\dd[3]{r}\sqrt{s_{\mathrm{x}}(\mathbf{r},t)^{2}+s_{\mathrm{y}}(\mathbf{r},t)^{2}}$. 
The predominant change (loss) of on-site magnetic moment during and after the action of the laser pulse is in the $z$ component, 
while the magnitude of the non-collinear part of the magnetization is only slightly affected.
\myFig{1}{1}{true}{-90}{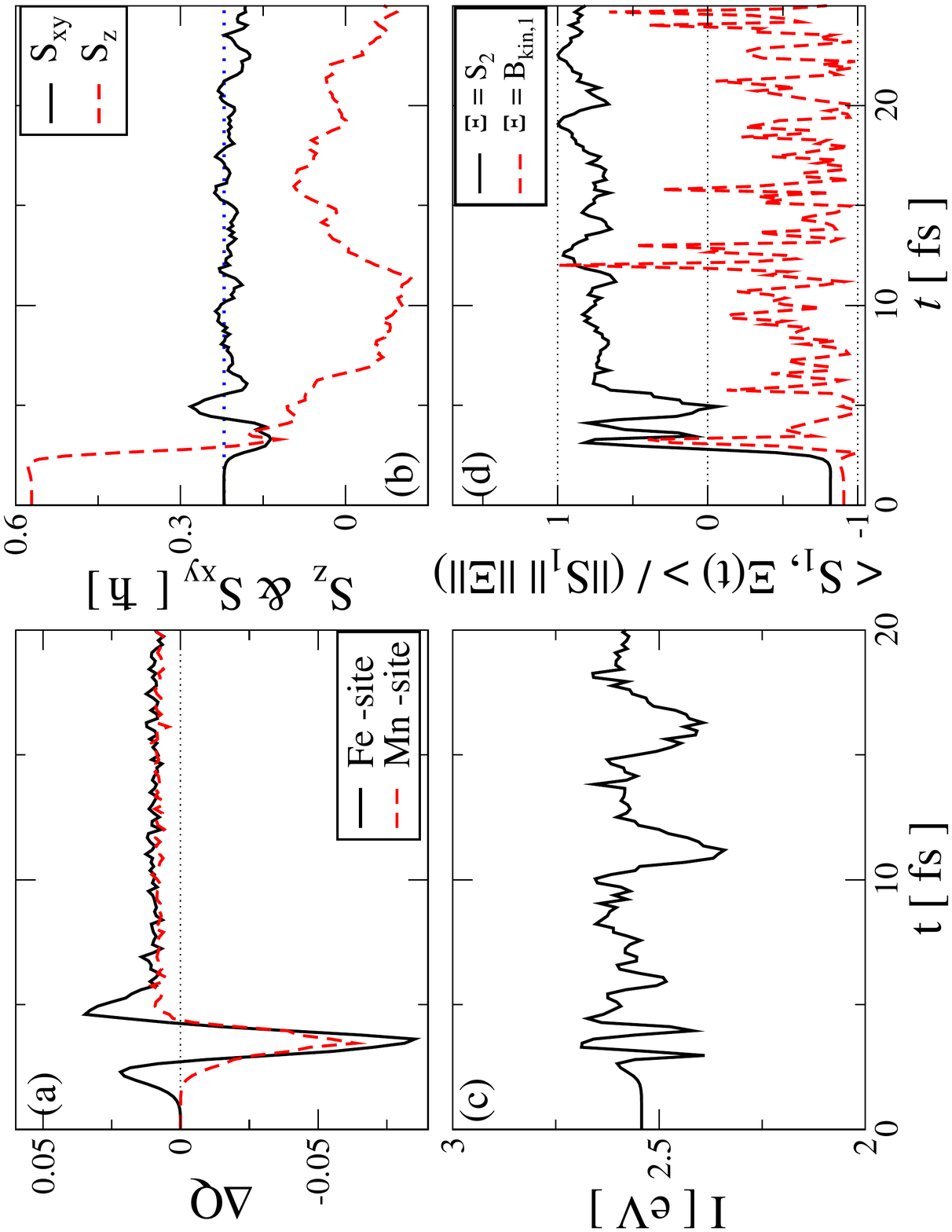}{(Color online) Time-averaged observables evaluated in spheres of radius $\SI{0.5}{\angstrom}$ around 
the F\lowercase{e} atomic site with the same type of pulses ($A=0.7$ eV/$\mathring{A}$): (a) the averaged temporal variation of the two 
on-site electronic charges $\Delta Q^{\mathrm{N}}(t) = Q^{\mathrm{N}}(t) - Q^{\mathrm{N}}(t=0)$, F\lowercase{e} site (black curve) 
and M\lowercase{n} site (red curve); (b) the averaged temporal variation of the magnetization along the $z$ axis, $S_z^{\mathrm{N}}(t)$, and 
the non-collinear component, $S_{xy}^{\mathrm{N}}(t)$, in correspondence of the F\lowercase{e} site; (c) dynamical Stoner exchange 
parameter $I(t)$ obtained by spatial integration, $I(t)=\frac{1}{\Norm{\mathcal{S}_{\mathrm{N}}^{\mathrm{R}}}}\int_{\mathcal{S}_{\mathrm{N}}^{\mathrm{R}}}\dd[3]{r} I(\mathbf{r},t)$, in correspondence of the F\lowercase{e} site (see Eq.~\ref{eq:btot}); (d) the averaged temporal variation of 
$F[\boldsymbol{\Xi}](t)=|\langle\mathbf{S}_{\mathrm{Fe}}(t), \boldsymbol{\Xi}(t)\rangle| / (\abs{\mathbf{S}_{\mathrm{Fe}}(t)}\cdot\abs{\boldsymbol{\Xi}(t)})$, 
where $\mathbf{S}_{\mathrm{Fe}}(t)$ is the on-site averaged magnetization computed within a radius $R = \SI{1.4}{\angstrom}$. 
Here $\boldsymbol{\Xi}(t)$ represents the spin calculated within the same radius over the M\lowercase{n} site (solid line) and the kinetic 
field $\mathbf{B}_{\mathrm{kin}}^{\mathrm{N}}(t)$ over the F\lowercase{e} site (dashed line). }{fig:02}
It is clear that, while the component $S_z^{\mathrm{N}}(t)$ collapses during the action of the electric field, after that it starts to oscillate 
around a new averaged value \footnote{Note that in Fig.~\ref{fig:02}(b) only the first period of the oscillation is shown.}. The behavior of the 
non-collinear component of the magnetization is instead different. After an initial bump during the action of the pulse, $S_{xy}^{\mathrm{N}}(t)$ 
returns to a value that is approximately equal to its initial one and then it remains constant for the rest of the time evolution. This 
kind of dynamics suggests the existence of different effective equations of motion for the two magnetization components. The reason for this 
very different dynamical behavior observed in the $S_z^{\mathrm{N}}$ and $S_{xy}^{\mathrm{N}}$ components lies in the fact that the Stoner 
excitations connect only up and down states along the spin quantization axis. In the initial ($t=0$) state the magnetic configuration of the system 
is, to a good level of accuracy, aligned along the $z$ axis. Hence, we would expect a pulse-driven Stoner-like excitation to mainly affect the 
$S_z^{\mathrm{N}}$ component and to a lesser degree the non-collinear $S_{xy}^{\mathrm{N}}$ magnetization.

In Fig.~\ref{fig:02} we introduce also a time-dependent effective Stoner parameter, $I(t)$. Typically within the DFT formalism $I$ is 
a measure of the drag in transferring charge density between the spin-up and spin-down bands of a solid. At the level of ground-state 
collinear spin DFT it is therefore commonly associated with the ratio between the exchange-correlation magnetic field,
$\mathbf{B}_{\mathrm{xc}}(\mathbf{r}, t)$, and the local value of the magnetization density. In order to generalize this concept to 
the case of a non-collinear magnetic system evolving in time we note that the effective local Hamiltonian, corresponding to the 
magnetic energy in Eq.~(\ref{eq:emag}), contains together with the exchange-correlation field also a second local contribution so 
that we can introduce the following effective local magnetic field,
\begin{equation}\label{eq:btot0}
 \mathbf{B}_{\mathrm{tot}}(\mathbf{r}, t) = \mathbf{B}_{\mathrm{xc}}(\mathbf{r}, t) + \mathbf{B}_{\mathrm{local}}(\mathbf{r}, t)\:.
\end{equation}
We use this expression to define a local Stoner vector, $\mathbf{I}(\mathbf{r}, t)$, parallel to $\mathbf{B}_{\mathrm{tot}}(\mathbf{r},t)$ 
and normalized with respect to the amplitude of the magnetization at each spatial point
\begin{equation}\label{eq:btot}
 \mathbf{B}_{\mathrm{tot}}(\mathbf{r}, t) = \mathbf{I}(\mathbf{r}, t)\cdot\abs{\mathbf{m}(\mathbf{r}, t)}\:.
\end{equation}
Hence, similarly to Eq.~(\ref{eq:btot0}), $\mathbf{I}(\mathbf{r}, t)$ can be identified from the sum of the two separate contributions 
originating from the local and the exchange-correlation field, 
$\mathbf{I}(\mathbf{r}, t)=\mathbf{I}_{\mathrm{xc}}(\mathbf{r}, t)+\mathbf{I}_{\mathrm{loc.}}(\mathbf{r}, t)$. In Fig.~\ref{fig:02}(c) we plot 
the module of the vector field $\mathbf{I}(\mathbf{r},t)$ integrated over a sphere centered at the F\lowercase{e} site. Its real time evolution 
within the first $20$~fs shows some oscillations activated by the action of the laser pulse. However, the overall change in the Stoner 
parameter during the evolution is not appreciable and $I(t)$ remains approximately constant throughout the entire dynamics and close 
to its initial value. This behaviour suggests that the initial dynamical change in the on-site $z$ component of the magnetization is mainly 
driven by the dissipation term $\nabla\cdot\mathcal{D}(\mathbf{r},t)$ and not due to a collapse of $I(t)$ which describes, instead, the resistance 
opposed by the band structure to inter-band transitions.

Fig.~\ref{fig:02}(d) shows the dynamical evolution of the function $F(t)$, which represents a measure of the normalized scalar product 
between the spin vector computed over the F\lowercase{e} site and the spin vector over the M\lowercase{n} site (solid line). During the 
action of the laser pulse $F(t)$ changes from an almost anti-ferromagnetic configuration (slightly non-collinear) to a ferromagnetic one, 
with the amount of spin misalignment being preserved during the process. We have seen that such an effect is determined by the Stoner 
excitations activated in the anti-ferromagnet by the action of the laser pulse. At longer times $F(t)$ oscillates around its new value and 
eventually approaches $1$, with the spin misalignment that is lifted out during the process. The dashed curve, instead, represents the 
evolution of $F[\boldsymbol{\Xi}](t)$, where $\boldsymbol{\Xi}(t)$ corresponds to the F\lowercase{e} on-site kinetic field. Within the first 
$5$~fs of the dynamical evolution $F(t)$ is characterized by strong fluctuations induced by the internal currents activated by the laser 
and its behaviour is similar to that described by the solid curve. However, after this initial phase, the evolution of $F(t)$ in the two cases 
appears quite different. Now, $F(t)$ strongly oscillates also after the action of the pulse inducing a torque on the magnetization vector. 
In practice, while the initial phase of the spin evolution is dominated by inter-band transitions activated by the action of the pulse, with 
consequent enhanced electronic hopping between the two atomic sites, after the action of the pulse the inter-band transitions are suppressed 
and the Kohn-Sham states evolve separately. The role played by the kinetic field becomes then more important inducing intra-band transitions 
with consequent spin relaxation over the two sites.

A further confirmation of these conclusions is provided by Fig.~\ref{fig:02b}(d), where we show a comparison between the local $z$ component 
of the magnetization, $S_z^{\mathrm{Fe}}(t)$, and its module $\abs{\mathbf{S}_{\mathrm{Fe}}(t)}$. During the action of the laser both the 
quantities decrease even if at different rates. After this first phase $S_z^{\mathrm{Fe}}(t)$ starts to oscillate around its new average value, 
while the module remains approximately constant. These two different dynamical behaviors may be explained in terms of initial inter-band 
transitions followed by an intra-band dynamical relaxation mechanism with the spin that is exchanged among the different components while 
its module remains constant.
 
Similarly we find in Fig.~\ref{fig:02b}(b) that the module of the exchange component $\abs{\mathbf{B}_{\mathrm{x}}^{\mathrm{Fe}}(t)}$, after 
the initial decay during the action of the laser remains approximately constant. In contrast the module of the local component of the kinetic field, 
$\abs{\mathbf{B}_{\mathrm{loc}}(t)}$, introduced in Eq.~(\ref{eq:emag}) appears, after the initial excitation, more oscillatory resembling the long-time 
dynamics of the $S_z^{\mathrm{Fe}}$ component. In Fig.~\ref{fig:02b}(a) we compare $\abs{\mathbf{B}_{\mathrm{loc}}^{\mathrm{Fe}}}$ under 
the application of the pulses shown in Fig.~\ref{fig:01}(b). In all the three cases this quantity is excited by the application of the pulse, but in the 
second phase of the dynamical evolution it collapses to a new lower value and it starts to oscillate around it. The application of different laser 
amplitudes does not seem to be reflected in a clear trend of the dynamical evolution of the local field. Finally we look at panel (c) where we plot 
the value of $\cos\theta$, with $\theta$ being the angle formed by the spin vector $\mathbf{S}_{\mathrm{Fe}}$ (black curve), or the kinetic field
$-\mathbf{B}_{\mathrm{kin}}^{\mathrm{Fe}}(t)$ (red curve) with the $z$ axis. This clearly shows, as we have already seen in Fig.~\ref{fig:02}(d), 
that after the application of the pulse the two vectors are highly non parallel with $\mathbf{B}_{\mathrm{kin}}$ playing a major role in the local 
dynamics of the spin vector.

\myFig{1}{1}{true}{-90}{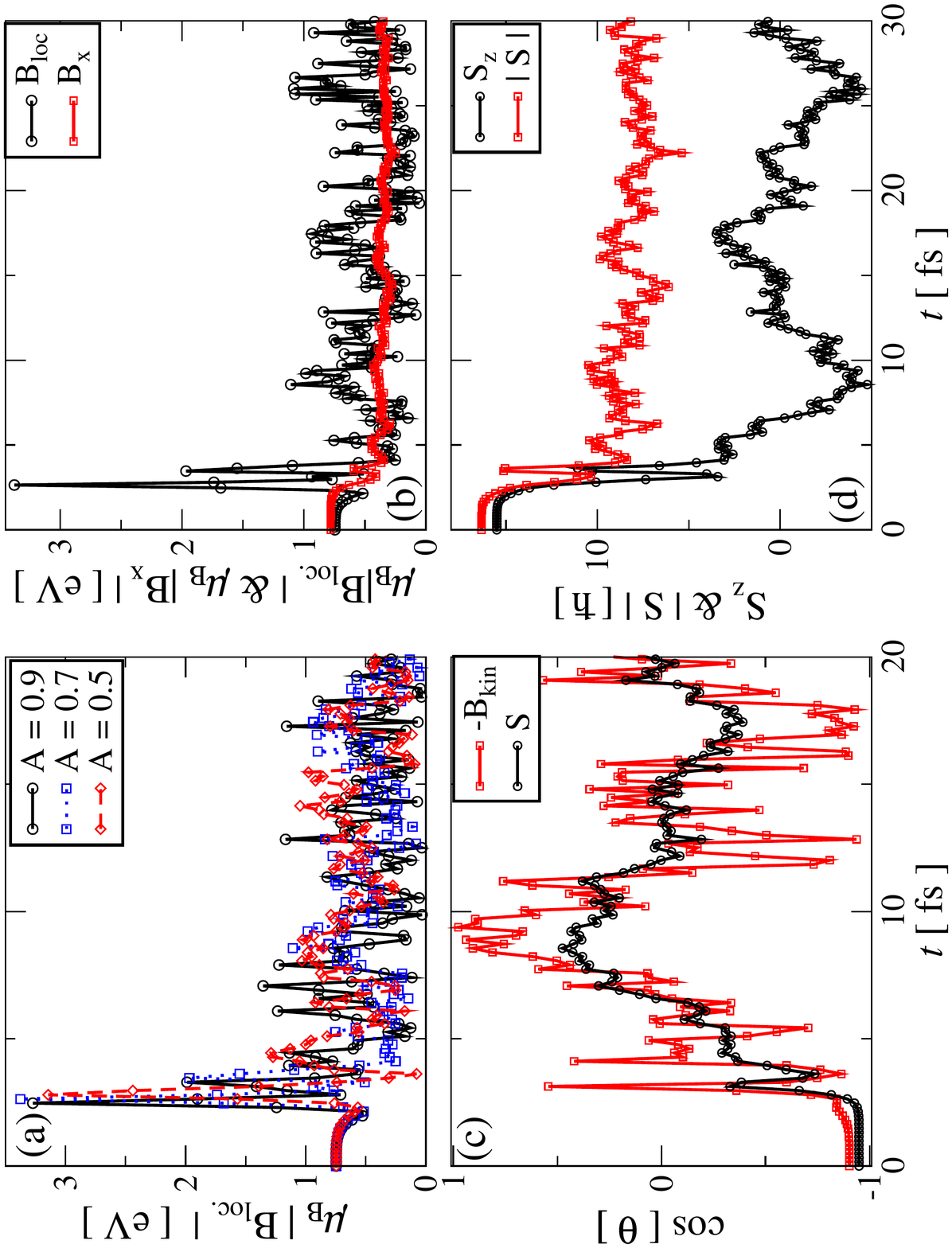}{(Color online) Time-averaged observables evaluated in spheres of radius $\SI{1.4}{\angstrom}$ around 
the atomic sites: (a) The averaged temporal evolution of $\abs{\mathbf{B}_{\mathrm{loc}}^{\mathrm{Fe}}(t)}$ for different pulse amplitudes 
$A = 0.5, 0.7, 0.9$~eV/\AA; (b) comparison of $\abs{\mathbf{B}_{\mathrm{loc}}^{\mathrm{Fe}}(t)}$ and 
$\abs{\mathbf{B}_{\mathrm{x}}^{\mathrm{Fe}}(t)}$ under a single pulse with amplitude $A = \SI{0.7}{eV / \angstrom}$; (c) averaged value 
of $\cos\theta$ during the time evolution, with $\theta$ angle formed with the $z$ axis by the $\mathbf{S}_{\mathrm{Fe}}(t)$ vector (black curve) 
and by the $\mathbf{B}_{\mathrm{kin}}^{\mathrm{Fe}}(t)$ vector (red curve), the pulse employed is the same of (b); (d) comparison between 
$\abs{\mathbf{S}_{\mathrm{Fe}}(t)}$ and $S_z^{\mathrm{Fe}}(t)$ during the temporal evolution, the pulse employed is the same of (b). }{fig:02b}

\myFig{1}{1}{true}{0}{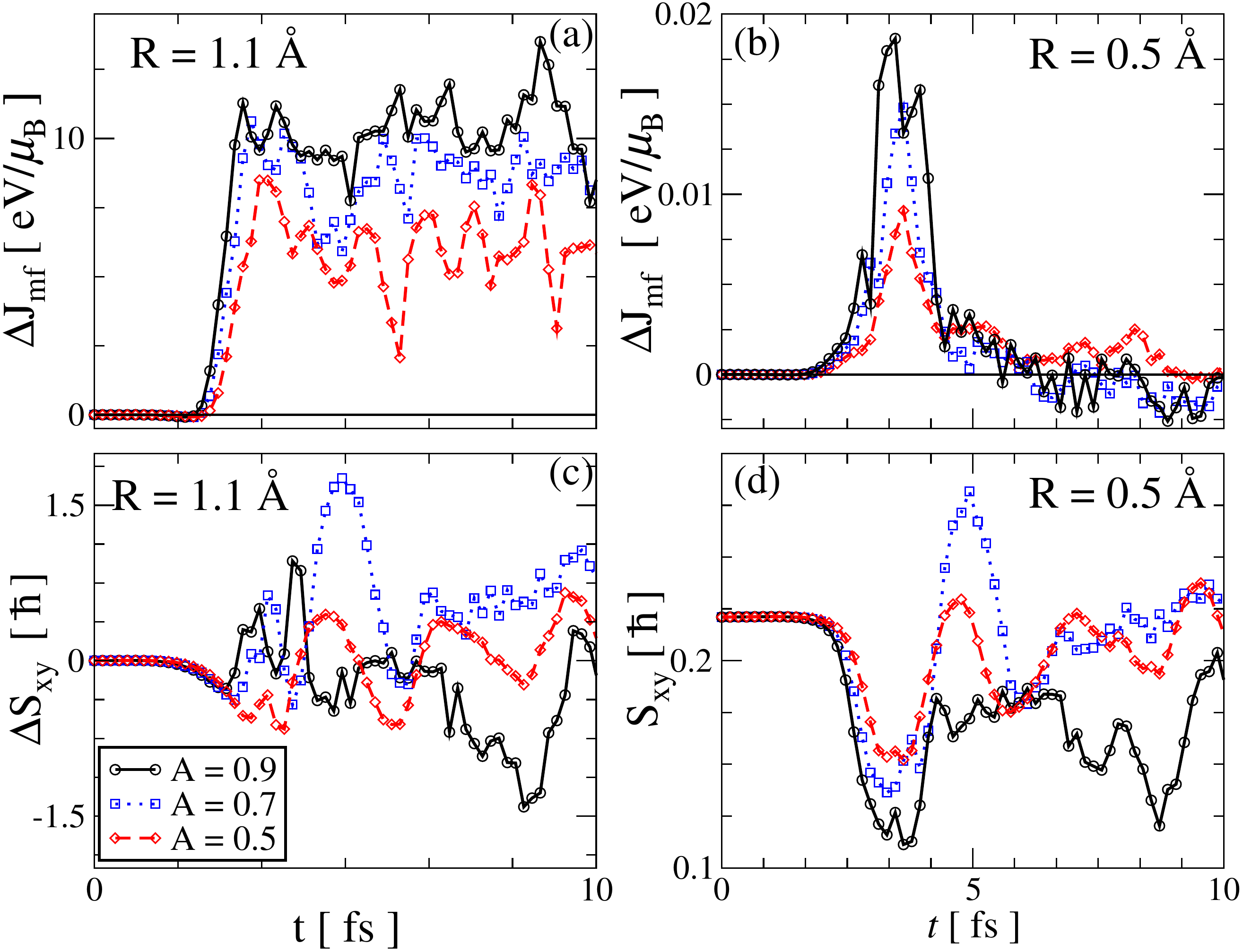}{(Color online) Exchange coupling evaluated inside spheres of different radii $R$ centered on the F\lowercase{e} 
atomic site and of $S_{\mathrm{xy}}^{\mathrm{Fe}}$ calculated inside the same spatial regions using three different type of pulses 
($A=0.7$, $A=0.9$, $A=0.5$)~eV/\AA. The dynamical exchange coupling here corresponds to the expression of 
Eq.~(\ref{eq:jcoupl}): (a) the averaged temporal variation of $\Delta J_{\textrm{mf}}(t) = J_{\textrm{mf}}(t)-J_{\textrm{mf}}(t=0)$ in the first 
$10$~fs; (b) the averaged temporal variation of 
$J_{\textrm{mf}}(t)$ inside a sphere of radius $R$; (c) the averaged temporal variation of $\Delta S_{\mathrm{xy}}^{\mathrm{Fe}}(t)$ 
obtained by spatial integration inside a sphere of radius $R$; (d) the averaged temporal variation of the same quantity 
$S_{\mathrm{xy}}^{\mathrm{Fe}}(t)$ evaluated inside a sphere of radius $R$. }{fig:03}

We can now focus our attention on the effective mean field term introduced in Eq.~(\ref{eq:spininter}) that has the form of a spin-spin 
interaction. This object, a non-local function of the magnetization vector density, can effectively be the source of spin waves in the 
dynamical evolution of the system. The temporal evolution of the dynamical exchange parameter $J_{\mathrm{mf}}(t)$ is presented in 
Fig.~\ref{fig:03} panels (a) and (b) and it appears to be strongly dependent on the laser-pulse excitation. In panel (b) it is shown that 
$J_{\mathrm{mf}}(t)$ follows the shape of the pulse. The value of the field is integrated within a sphere of small radius 
($R = \SI{0.5}{\angstrom}$) around the F\lowercase{e} site, therefore, higher pulse amplitudes excite more the electronic system with 
consequent higher modification of $J_{\mathrm{mf}}$. The quantity sharply increases from its initial value and then returns close to its 
ground-state magnitude on the time scale of the pulse disappearence. The maximum amplitude of $J_{\mathrm{mf}}$ also scales 
systematically with the amplitude of the laser pulse, i.e. it increases for the more intensive pulses. The trend of $J_{\mathrm{mf}}(t)$ 
shown in panel (a) is very similar, the quantity is computed within a sphere of larger radius. While during the application of the pulse 
$J_{\mathrm{mf}}(t)$ looks strongly affected, and its growth rate scales proportionally to the pulse amplitude, after that the exchange 
coupling stabilizes around an average value, different in the three cases. This reflects the different amount of energy injected into the 
system by the three pulses.

In comparison, Figure~\ref{fig:03} panels (c) and (d) show the evolution of the non-collinear spin function, $S_{xy}^{\mathrm{Fe}}(t)$, 
for the same set of simulations with increasing laser-pulse intensity. We find clear similarities during the pulse-coherent stage of the 
dynamics. Both $J_{\textrm{mf}}(t)$ and $S_{xy}^{\mathrm{Fe}}(t)$ follow the pulse, and their amplitudes vary with the pulse intensity. 
At times longer than the pulse duration the dynamics of the two objects, however, is completely different. This difference stems from 
the fact that the non-collinear spin component $S_{xy}^{\mathrm{N}}(t)$ is driven by two torques, as described by Eq.~(\ref{eq:spininter}), 
and only a part of the second torque is related to $J_{\textrm{mf}}$. The dynamical exchange coupling shown in Fig.~\ref{fig:03}(b) 
is characterized by a large variation during the action of the laser pulse and it could, at least in principle, activate an out-of-equilibrium 
dynamics involving the non-collinear components of the two atomic spins. We will see now that this is indeed the case. 

\myFig{1}{1}{true}{-90}{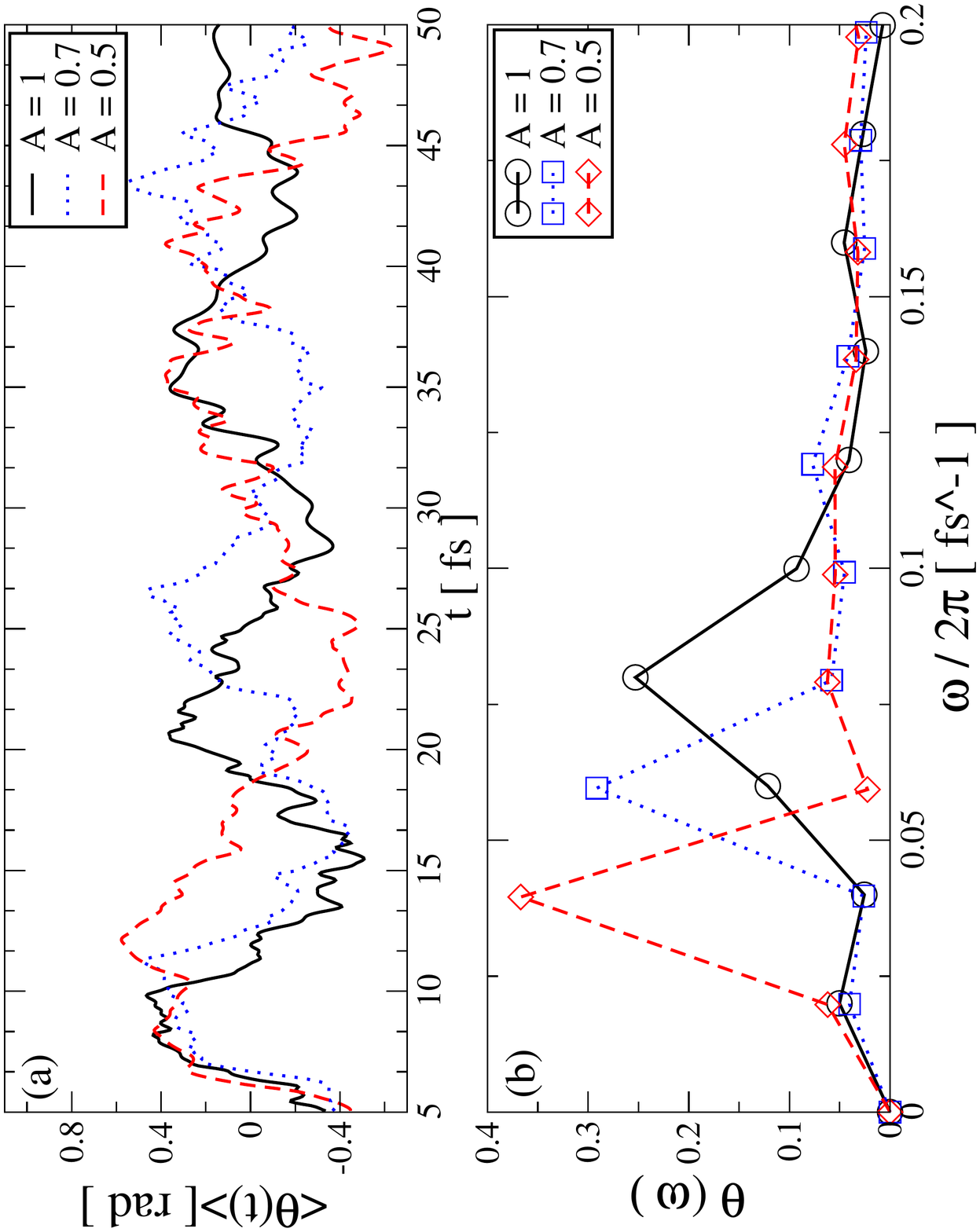}{(Color online) Time-averaged observables evaluated in spheres of radius $\SI{0.5}{\angstrom}$ 
around the F\lowercase{e} atomic site with three different type of pulses ($A=1.0, 0.7, 0.5$~eV/\AA): (a) the temporal variation of 
the angle $\theta(t)=\arccos\big(S_z^{\mathrm{Fe}}(t) / \abs{\mathbf{S}^{\mathrm{Fe}}}\big)$ with respect to its temporal averaged value 
$\langle\theta(t)\rangle = \frac{\sum_{n=1}^{N}\theta(t_{n})}{N}$; (b) the Fourier transform $\mathfrak{F}(\theta)(\omega)$ where 
$\theta(t)$ is the angle previously defined. }{fig:04}

Further evidence of the validity of this argument for the spin-spin exchange are provided in Fig.~\ref{fig:04}(b), where we present the 
Fourier transform of the angle $\theta(t)$ formed by the spin vector, $\mathbf{S}^\mathrm{Fe}(t)$, on the F\lowercase{e} site [see Eq. (\ref{eq:Mloc})] 
with the $z$ axis. In panel (a) we present the corresponding temporal evolution of $\theta(t)$ by measuring the on-site spin misalignment (F\lowercase{e} atom). As before we compare the results from 
the three different simulations with increasing pulse amplitudes. Focusing on the lowest part of the spectrum, we observe that the lowest 
frequency peak in the spectrum blue-shifts with increasing the pulse amplitude. Although our resolution is limited by the length of the 
numerically-stable time integration and we only observe one or two periods of the lowest frequency mode, it is clear that the laser pulse 
amplitude affects directly the energy of that spin-wave mode. This correlates with our previous observation of strongly pulse-intensity 
dependent effective exchange interaction $J_{\textrm{mf}}$.

In a summary, it is clear that the laser excites directly only the electronic system and this is propagated to the spin system through the 
consequent formation of Stoner excitations and transfer of magnetization between the two sites. However, at the same time the excitations 
in the electronic sub-system contribute also to the ultrafast modification of the effective inter-site EDEI $J_{\textrm{mf}}(t)$ 
depicted in Fig.~\ref{fig:03}(a). Certainly, the degree of laser-induced modification of this quantity is proportional to the amount of energy 
injected into the system by the pulse, namely to the amplitude of the applied external field. The physical interpretation of $J_{\textrm{mf}}(t)$ 
as a Heisenberg-like exchange parameter is further validated by the observation that the lower energy spin-wave modes follow an analogous 
dependence on the excitation magnitude.

\section{Conclusions}\label{sect:4}
In conclusion, starting from the hydrodynamical formulation of the spin dynamics in the ALDA, we introduce an out-of-equilibrium non-local 
spin-spin interaction term and define an effective out-of-equilibrium Heisenberg-like exchange coupling. We evaluate the latter through 
TD-SDFT calculations by applying ultrafast external electric pulses (of duration of about 5~fs and amplitudes ranging between 
$A=0.5$~eV/{\AA} and $A=1$~eV/\AA) to {\it fcc} F\lowercase{e}M\lowercase{n}, which has a frustrated anti-ferromagnetic 
ground-state. These simulations show that the observed on-site demagnetization can be attributed mainly to a Stoner-like excitation 
activated by the action of the laser pulse. The local dynamics, at longer time, appears to be driven by an inter-site exchange coupling, 
$J_{\mathrm{mf}}(t)$. After a strong non-adiabatic modification during the action of the pulse the EDEI acquires a new value that remains 
approximately constant for the rest of the evolution. The analysis of the Fourier spectrum of the angle $\theta(t)$ formed by the on-site spin 
direction with the $z$ axis suggests that the laser pulse can activate spin wave excitations with a frequency that grows with the amplitude 
of the applied pulse and correlates also with the new value acquired by the effective exchange coupling $J_{\mathrm{mf}}$.

\acknowledgments
This work has been funded by the European Commission project CRONOS (Grant No. 280879) and by Science Foundation Ireland 
(Grant No. 14/IA/2624). We also gratefully acknowledge the DJEI/DES/SFI/HEA Irish Centre for High-End Computing (ICHEC) for the 
provision of computational facilities and support and  the Trinity Centre for High Performance Computing for technical support.


\begin{thebibliography}{hbp}

\bibliography{research}

\bibitem{Kat00} M.I.~Katsnelson and A.I.~Lichtenstein, Phys. Rev. B \textbf{61}, 8906 (2000).

\bibitem{Kat04} M.I.~Katsnelson and A.I.~Lichtenstein, J. Phys.: Condens. Matter \textbf{16}, 7439 (2004).

\bibitem{Loun10} S.~Lounis and P.H.~Dederichs, Phys. Rev. B \textbf{82}, 180404(R) (2010).

\bibitem{EXCH3} A.~Jacobsson, B.~Sanyal, M,~Le\u{z}ai\'{c} and S.~Bl\"{u}gel, Phys. Rev. B \textbf{88}, 134427 (2013).

\bibitem{MFT} A.I.~Liechtenstein, M.I.~Katsnelson, V.P.~Antropov and V.A.~Gubanov, J. Magn. Magn. Mater. \textbf{67}, 65 (1987).

\bibitem{MFT2} V.P.~Antropov, J. Magn. Magn. Mater. \textbf{262}, 192 (2003).

\bibitem{EXCH} P.~Bruno, Phys. Rev. Lett. \textbf{90}, 087205 (2003).

\bibitem{EXCH2} A.~Szilva, M.~Costa, A.~Bergman, L.~Szunyogh, L.~Nordstr\"om and O.~Eriksson, Phys. Rev. Lett. \textbf{111}, 127204 (2013).

\bibitem{Liu89} K.L.~Liu and S.H.~Vosko, Can. J. Phys. \textbf{67}, 1015 (1989).

\bibitem{TDDFT} M.A.L~Marques and E.K.U.~Gross in {\it A Primer in Density Functional Theory}, edited by C.~Fiolhais, F.~Noqueira and M.~Marques, Lecture Notes in Physics Vol. 620 (Springer-Verlag, Berlin, 2003)

\bibitem{Gross} E.K.U.~Gross and W.~Kohn, Adv. Quantum Chem. {\bf 21}, 255 (1990).

\bibitem{Stamen}M.~Stamenova and S.~Sanvito, Phys. Rev. B {\bf 88}, 104423 (2013).

\bibitem{Peral15} J.E.~Peralta, O.~Hod and G.E.~Scuseria, J. Chem. Theory Comput. \textbf{11}, 3661 (2015).

\bibitem{Kim} A.V.~Kimel, A.~Kirilyuk and Th.~Rasing Laser \& photon. Rev. \textbf{1}, 275 (2007).

\bibitem{BEAU} E.~Beaurepaire, J.-C.~Merle, A.~Daunois and J.-Y.~Bigot, Phys. Rev. Lett. \textbf{76}, 4250 (1996).

\bibitem{Carpene} E.~Carpene, H.~Hedayat, F.~Boschini and C.~Dallera, Phys. Rev. B \textbf{91}, 174414 (2015).

\bibitem{EXCT2} G.P.~Zhang, M.~Gu and S.X.~Wu, J. Phys.:Condens. Matter {\bf 26}, 376001 (2014).

\bibitem{Sec13} A.~Secchi et Al., Annals of Physics \textbf{333}, 221 (2013).

\bibitem{dynEXCH} B.Y.~Mueller, A.~Baral, S.~Vollmar, M.~Cinchetti, M.~Aeschlimann, H.C.~Schneider and B.~Rethfeld, Phys. Rev. Lett. \textbf{111}, 167204 (2013).

\bibitem{Elliot16} P.~Elliott, T.~M\"uller, J.K.~Dewhurst, S.~Sharma and E.K.U.~Gross, Sci. Rep. \textbf{6}, 38911 (2016).

\bibitem{Zhang15} G.P.~Zhang, M.S.~Si, Y.H.~Bai and T.F.~George, J. Phys.: Condens. Matter \textbf{27}, 206003 (2015).

\bibitem{Taka55} T.~Takabayasi, Prog. Theor. Phys. \textbf{14}, 283 (1955).

\bibitem{Antr1} V.P.~Antropov, J. Appl. Phys. \textbf{97}, 10A704 (2005).

\bibitem{Antr2} M.I.~Katsnelson and V.P.~Antropov, Phys. Rev. B \textbf{67}, 140406(R) (2003).

\bibitem{Antr3} V.~Antropov, M.I.~Katsnelson, B.N.~Harmon, M.~van Schilfgaarde and D.~Kusnezov, Phys. Rev. B {\bf 54}, 1019 (1996).

\bibitem{me} J.~Simoni, M.~Stamenova and S.~Sanvito, Phys. Rev. B \textbf{95}, 024412 (2017).

\bibitem{spindyn} K.~Capelle, G.~Vignale and B.L.~Gy\"orffy, Phys. Rev. Lett. \textbf{87}, 206403 (2001).

\bibitem{femn1} M.~Ekholm and I.A.~Abrikosov, Phys. Rev. B {\bf 84}, 104423 (2011).

\bibitem{femn2} C.~Grazioli, D.~Alf\'e, S.R.~Krishnakumar, S.S.~Gupta et Al., Phys. Rev. Lett. {\bf 95}, 117201 (2005).

\bibitem{alda} K.~Yabana and G.F.~Bertsch, Phys. Rev. B {\bf 54}, 4484 (1996).

\bibitem{Wang92} J.~Zhu, X.W.~Wang and S.G.~Louie, Phys. Rev. B {\textbf 45}, 8887 (1992). 

\bibitem{Octop1} A.~Castro, H.~Appel, M.~Oliveira, C.A.~Rozzi, et al., Phys. Stat. Sol. B \textbf{243}, 2465 (2006).

\bibitem{mrpp} C.L.~Reis, J.M.~Pacheco and J.L.~Martins, Phys. Rev. B {\bf 68}, 155111 (2003).

\bibitem{APE} M.~J.~T.~Oliveira and F.~Nogueira, Comp. Phys. Comm. {\textbf 178}, 524 (2008).

\bibitem{APE2} M.A.L.~Marques, M.J.T.~Oliveira and T.~Burnus, Comp. Phys. Comm., {\textbf 183}, 2272 (2012).

\end{thebibliography}
\end{document}